\documentstyle[12pt]{article}
 \newcommand \be {\begin{equation}}
\newcommand \bea {\begin{eqnarray} \nonumber }
\newcommand \ee {\end{equation}}
\newcommand \eea {\end{eqnarray}}
 \newcommand \eps {\epsilon}
 
\newcommand \de {\delta}

 \newcommand \al {\alpha}

\newcommand{\bi}{\bibitem}

\newcommand{\nn}{\nonumber\\}
 \def\(({\left(}
 \def\)){\right)}
\def\[[{\left[}
\def\]]{\right]}
\def\crho{{\hat \rho}}
\def\tg{{\tilde g}}
\def\th{{\tilde h}}
\def\tW{{\tilde W}}
\def\btW{{\bf \tilde W}}

\def\bth{{\bf \tilde h}}
\def\bW1{{\bf  W_1}}
\def\bh1{{\bf  h_1}}
\begin{document}

\title{A tentative replica study of the glass transition} 

\author{ Marc M\'ezard$^1$ and Giorgio Parisi$^2$}

\date{\it $^1$ 
 Laboratoire de Physique Th\'eorique de l' Ecole Normale Sup\'erieure
 \footnote {Unit\'e propre du CNRS,  associ\'ee
 \`a\ l'Ecole
 Normale Sup\'erieure et \`a\ l'Universit\'e de Paris Sud} , 
\\ 24 rue
 Lhomond, 75231 Paris Cedex 05, France \\ 
\hbox{  }  \hfill \break
$^2$ 
Dipartimento di Fisica,
Universit\`a {\sl La  Sapienza}\\ INFN Sezione di Roma I \\ Piazzale
Aldo Moro, Roma 00187, Italy }
\maketitle
\begin{abstract}
 In this paper we propose a method  to study quantitatively 
the glass transition
in a system of interacting particles. In spite of the absence
of any quenched disorder, we introduce a replicated version of
the hypernetted chain equations. The solution of these equations,
for hard or soft spheres, signals a transition to the glass
phase. However the predicted value of the energy and specific
heat in the glass phase are wrong, calling for
an improvement of this method.
 \end{abstract}

\vskip 0.5cm

LPTENS preprint 96/12.

\vskip 0.5cm

\noindent Electronic addresses : mezard@physique.ens.fr, parisi@roma1.infn.it

\newpage
%123456789%123456789%123456789%123456789%123456789%123456789%123456789%123456789
\section{Introduction}
The long lasting studies of the glass transition, in spite of
recent progress \cite{Science}, have not yet produced a
completely convincing theory deriving the low temperature behavior from
a microscopic Hamiltonian of interacting particles.

It has been proposed that near this transition real glasses could behave in the 
same way as some disordered systems which are described by a Hamiltonian with 
quenched randomness \cite{KiThi,KiThiWo,Parglass,FrHe,BCKM}.  This proposal is 
supported the existence of a wide class of mean field systems, with fixed non 
random Hamiltonians which show a typical glassy behaviour
\cite{BM,MPR,FrHe,SI3,Joseph,JosephPar,MPR_FF,CKMP}.
Although they depend on one or a few coupling constants, these systems behave as 
if they were typical samples of a class of systems with quenched randomness 
(apart from the possible existence of a crystal phase\cite{MPR}).

The relevant disordered systems found so far have the following behaviour at the 
mean field level.  There are two phase transitions: a usual static transition at 
low temperature ($T_R$), where the replica symmetry is broken and the specific 
heat is discontinuous, and a dynamic transition at a higher temperature ($T_D$), 
where some time persistent correlations set in.  In the region where $T<T_D$ the 
infinite system, quenched from some high energy configuration, gets trapped into 
metastable states, such that its energy density is larger than the equilibrium 
value.  In this dynamics the metastability appears at $T_R$ and the 
thermodynamic transition, which is present at a lower temperature, is not 
accessible\cite{KiThi,CriHorSom,CuKu,FraPar,BaBuMez}.  Beyond the mean field 
approximation one finds that the equilibration time is not divergent at $T_D$, 
but it becomes exponentially large in the region $T_C < T <T_D$.  It has been 
argued that in some spin model the equilibration time diverges (as $\exp (A 
|T-T_C|^{-\gamma)}$, $\gamma=d-1$ in $d$ dimensions \cite{KiThiWo,Parglass}.  
This problem will not be studied here, we shall stay at a mean field level.

In this paper we want to present a progress report on an analytical study of the 
glass transition for a real three dimensional model of interacting particles.

\section{Breaking the replica symmetry}

We consider a system of $N$ interacting particle in a volume $V$. 
We study the infinite volume limit in which  $N \to \infty$ at fixed 
$\rho \equiv N/V$. The Hamiltonian is given by:
\begin{equation}
	H(x)=\sum_{i < k} U(x_i-x_k)
	\label{Hamiltonian}
\end{equation}

We will consider two forms of the potential among particles:
  The hard sphere case where $U(x)=\infty$ for $r\equiv|x|<1$ and $U(x)=0$ 
	for $r>1$, and
a soft sphere case $U(x) =r^{-12}$.

In both cases, if the system is cooled from  high 
 to  low temperature fast enough, crystalization is 
inhibited and the system undergoes a glass transition.
 For hard spheres this transition occurs at a density of about 
$1.15 $ independently from the temperature, while the crystallisation
would occur for an equilibrium system at density 
around $.95$.
 In the case of soft spheres thermodynamic quantities are functions 
of the dimensionless density 
$\gamma\equiv \rho T^{(-1/4)}$ and the glass
transition is 
located around 
$\gamma\simeq1.6$, while the freezing value of $\gamma$ is
 around $\gamma \simeq 1.15$ \cite{hiw}.

It is possible to consider  $n$ replicas of the same system and to 
study the total partition function 
\be
Z= \int \prod_{i,a} dx_i^a \exp-\beta\((  \sum_a H(x^a) + \eps \sum_{a,b}
\sum_{i \ne k} \phi_{a,b}(x^a_i-x^b_k)\)) \ .
\ee
The detailed form of the (attractive) potential $\phi_{a,b}$ is not relevant here 
because we will eventually set $\eps=0$.

Replica symmetry breaking corresponds to the case in which (in the infinite 
system) the correlation among replicas does not vanish in the limit $\eps \to 
0$.  The physical meaning of this breaking is clear: at high density, if several 
metastable states contribute to the Boltzmann measure, different replicas may be 
frozen in the same state.  Indeed we can propose the following picture of the 
glass phase. 

There are many equilibrium states, which we label by an index $\al$ (modulo 
rotations and translations).  These states are identified by the density 
$<\crho(x)>_\al\equiv \rho_\al(x)$ which depends on $x$ in each state.  In this 
paper we denote by $\crho(x)$ the density operator $\sum_j
\delta(x-x_j)$ and similarly by $\crho_a(x)$ the density operator in replica 
$a$.  The brackets stand for thermal expectation values.  Each state $\al$ 
appears in the Boltzmann measure with a weight $w_\al$.  We assume that there 
are many states, and the quantities which we can access are averaged over the 
states.  Since the one point average $\sum_\al w_\al \rho_\al(x)= \rho$ is 
uniform, we need to study the two point correlations like:
\bea
\tg(x,y)&=&{1 \over \rho^2} \sum_\al w_\al
 \sum_{i \neq j} <\de(x_i-x) \de(x_j-y)>_\al
\nn 
g_1(x,y)&=&{1 \over \rho^2 \sum_\al w_\al^2}
 \sum_\al w_\al^2 \sum_{i \neq j} <\de(x_i-x)>_\al < \de(x_j-y)>_\al \ ,
\label{gdef}
\eea
 As for the densities in 
two different states, $\al \neq \beta$, we assume that they are uncorrelated.

In replica space one could thus expect the correlations to be, for $a \neq b$:
$
<\crho_a(x) \crho_b(y)> = w_2 \rho_\al(x) \rho_\al(y)= w_2 \rho^2 g_1(x,y) 
$,
where
we use the notation
\be
w_k=<\sum_\al w_\al^k> =\int dw  \ P(w) \ w^k.
\ee
Notice that we are describing the quantities $w$ by their probability 
distribution $P(w)$.  This description may look strange for such a deterministic 
problem (the Hamiltonian is not random).  It is clear that in a finite volume, 
at fixed $N$, the $w$'s are given numbers.  However they may strongly depend on 
$N$ so that if we average over a small window in $N$ around $\overline{N}$ and 
send $\overline{N} \to \infty$ we induce a smooth probability distribution for 
these variables.  In this formulation one finds the unpleasant feature that the 
correlation function of more replicas do not factorize when the distance goes 
infinity.  For instance, the four point function
\be
<\crho_a(x) \crho_b(y) \crho_c(z) \crho_d(t)>=
 \sum_{\al} w^4_\al \rho_\al(x) \rho_\al(y)  \rho_\al(z) \rho_\al(t)
\ee
goes  to 
$w_4 g_1(x,y) g_1(z,t)$
when $x-y=O(1)$, $z-t=O(1)$ and $x-z \to \infty$, which is 
different from the factorized form  $(w_2)^2 g_1(x,y) g_1(z,t)$. 
 
The non factorization of the correlation function is deeply linked to the 
probability distribution of the variables $w$.  This crucial ingredient of the 
description is usually dealt with by using the replica symmetry breaking 
formalism.  This means that the correlation functions (e.g.  $<\crho_a(x) 
\crho_b(y)>$) are no more symmetric under the permutation of the replicas, but 
they have the advantage of being clustering \footnote{We thank Francesco Guerra 
for discussions on this point.}.  This formalism provides and automatic 
bookkeeping of all complications which would arise from the existence of many 
states.  If one assumes a given form for replica symmetry breaking, it 
corresponds to a given form for the probability distribution of the $w$.  In the 
simplest case, called one step breaking, one divides the $n$ replicas in $n/m$ 
groups of $m$ replicas and one assumes the following structure of correlation 
functions:
\bea
<\crho_a(x) \crho_a(y)> &=& \rho^2 \tg(x-y) \ , \nn
  <\crho_a(x) \crho_b(y)> &=& \rho^2 g_1(x-y)  \mbox{ for $a \ne b$ in the same 
group} \ ,\nn
<\crho_a(x) \crho_b(y)> &=&\rho^2  \mbox{ for $a$ and  $b$ in  different 
groups.}\  .
\label{ONESTEP}
\eea

In the  physically relevant case where $m<1$, this form of 
the correlations corresponds to assume that  
$
w_\al =  \exp( -\beta f_\al) /[ \sum_{\al'} \exp( -\beta f_{\al'})]
$
where the $f_\al$ are negative quantities 
extracted with a probability distribution 
 $\beta m \exp ( \beta m f)$. The details of this construction are 
 described extensively in the literature \cite{MPV}.

May be the reader has some familiarity with the replica formalism and he 
is waiting here for the canonical sentence {\it we send $n$ to zero at 
the end of the computation}. However this is not a random system and 
there is nothing to average (apart from the value of $N$), 
there is no need to send $n$ to zero. 
Indeed it turns out that in this scheme, the free energy density,
$-\ln(Z)/(\beta N n)$, is independent 
of $n$ as soon as $<\crho_a(x) \crho_b(y)> - <\crho_a(x)>
 < \crho_b(y)> =0$   for $a$ and  $b$ in  different 
groups. In some sense $n$ is infinite, because we assume that $m$ divides 
$n$ for any $m$.

In this scheme the free energy depends on $m$, which should be considered as a 
variational parameter and the free energy must be maximized (not minimized!) as 
usual in the replica approach.  For $m=1$ we recover the replica symmetric 
result.  The static glass transition temperature $T_R$ is characterized by the 
existence of a maximum of the free energy at $m(T)<1$, where $m(T_R)=1$.  On the 
other hand, the appearance of metastable states which will trap the dynamical 
evolution of the system is signalled by the existence of a non trivial $g_1(x)$ 
in the limit $m \to 1$ \cite{FraPar,Mon}.

The programme is thus clear.  The real difficulty consists now in implementing 
it, i.e.  in computing the properties of our replicated system of interacting 
particles, allowing for replica symmetry breaking.  A first step in this 
direction will be proposed in the next section.

\section{The replicated hypernetted chain approximation}

The hypernetted chain (HNC) approximation consists in considering only a given 
class of diagrams in the virial expansion \cite{HanMc}.  It gives a reasonable 
account of the liquid phase, and it has also been used for studying the first 
order transition to the crystal phase \cite{RamYus}.  We will consider here this 
approximation because it has the advantage of having a simple variational 
formulation.  In the liquid phase, where the density is constant, the usual HNC 
equation (for the non replicated system) can be written as
\be
g(x) = \exp \left(-\beta U(x) + W(x)\right),
\ee
where:
\bea
\rho^2 g(x) &=& \rho^2 (1+h(x))= <\crho(x)\crho(0)>- \rho \de(x),\\
W(x)&\equiv& 
 \int {d^dp \over (2 \pi)^d} \ e^{-ipx} 
{\rho {\bf h}(p)^2 \over 1+\rho {\bf h}(p)}\ , 
\label{HNC}
\eea
and we denote by ${\bf h}(p)$  the Fourier transform of $h(x)$.

This equation can be derived by minimizing with respect to 
$h(x)$ the following free energy per 
unit volume, in the space of functions of $|x|$:
\be
2
\beta F = \int d^dx \ \rho^2 g(x)[\ln (g(x))-1 +\beta U(x)] +
 \int {d^dq \over (2 \pi)^d} L(\rho {\bf h}(q)),
\ee
where $L(x) \equiv -\ln(1+x)+x-x^2/2$.

In this note we propose a bold generalization of the HNC equations for $n$ 
replicas.  The replicated free energy is now
 \be
2
\beta F = \rho^2 \int d^dx \sum_{a,b} g_{ab}(x)
[\ln (g_{ab}(x))-1 -\beta ( U(x) \delta_{a,b}+\eps  \phi_{ab}) ] +\mbox{Tr}
L(\rho {\bf h}),
\ee
where  ${\bf h}$ is now an operator both in $x$ space and in 
replica space.

If in the limit $\eps \to 0$ one finds that at large enough densities $g_{ab}$ 
is non zero off the diagonal, replica symmetry is broken.  In the case where 
$g_{ab}$ is of the form shown in eq.  (\ref{ONESTEP}), this equation can be used 
to compute the properties of the correlation function in the glassy phase.  This 
approach amounts to a study of the density modulations in the glass phase at the 
level of the two point function.  In the glass phase $\rho_\alpha(x)$ becomes 
space dependent.  However, as argued in the introduction, the necessity of 
averaging over the states $\al$ forces us to study this $x$ dependence at the 
level of the two point correlations: We keep here to the definition of $\tg$ 
through $
\sum_\al w_\al <\crho(x) \crho(y)>_\al = 
\de(x-y) \sum_\al w_\al \rho_\al(x) + \rho^2 \tg(x,y)
$, while the second term should
be written more properly as $\rho_\al(x) g_\al(x,y) \rho_\al(y)$.
So our correlation $\tg$ reflects the  structure of $\rho_\al(x)$ 
as a sum of peaks of unit weights, smoothed by the average over states.

Within the one step breaking scheme (\ref{ONESTEP}), the free energy is:
\bea
\lim_{n \to 0}{
2
\beta F \over n \rho^2}=  \int d^d x \{\tg(x)[\ln(\tg(x))-1
+\beta U(x) ] -(1-m) g_1(x)[\ln(g_1(x))-1] \} 
\\
-\int {d^d q \over (2\pi)^d} \ 
 \((
{\th(q)^2 \over 2}-(1-m){h_1(q)^2 \over 2}-{\th(q) \over \rho} \right.
\nn
\left.
+{1 \over m \rho^2} \ln[1+\rho \th(q)-(1-m) \rho h_1(q)]
-{1-m \over m \rho^2} \ln[1+\rho \th(q)- \rho h_1(q)]
\))
\eea

The static transition is identified as
 the temperature (or density) at which
there exists a non trivial solution to the replicated HNC equations
$ \partial F/ \partial \tg(x)=0$, $ \partial F/ \partial g_1(x)=0$
and $ \partial F/ \partial m=0$, for $ m \in [0,1]$
(in fact we must minimize the free energy with respect
 to $\tg$, but maximize 
with respect to $g_1$ and $m$). 
The equations to be solved for the statics are thus:
\bea
\tg(x)=\exp\((-\beta U(x)+\tW(x)\)),\ \ g_1(x)=\exp\((W_1(x)\))
\\
\int d^dx  \ g_1(x) [\ln(g_1(x))-1]=\int {d^dq \over (2 \pi)^d}
\(( 
{\bh1(q)^2 \over 2} 
+ {1 \over m \rho}{\bh1(q) \over 1+\rho \bth(q)-(1-m) \rho \bh1(q)}
\right. \\ \left.
-{1 \over m^2 \rho^2}
\ln{ 1+\rho \bth(q)-(1-m) \rho \bh1(q) \over 1+\rho \bth(q)- \rho \bh1(q)}
\))
\label{HNCrepstat}
\eea
where the Fourier transforms $\btW(q)$ and $\bW1(q)$ of $\tW(x)$ and
$W_1(x)$ satisfy:
\bea
 \btW(q)-(1-m) \bW1(q)&=& \rho { (\bth(q)-(1-m)\bh1(q))^2 \over
1+\rho(\bth(q)-(1-m)\bh1(q))}
\\
\bW1(q)&=&{1 \over m} \((
{ \rho (\bth(q)-(1-m)\bh1(q))^2 \over 1+\rho(\bth(q)-(1-m)\bh1(q))}
-{ \rho (\bth(q)-\bh1(q))^2 \over 1+\rho(\bth(q)-\bh1(q))} \))
\label {Wdef}
\eea

The dynamical transition may be characterized as the highest 
temperature at which there is a non trivial solution of the
stationarity equations $ \partial F/ \partial \tg(x)=0$ and
 $ \partial F/ \partial g_1(x)=0$ at $m=1^-$. The corresponding
equations are obtained by substituting $m \to 1$ in the
first two equations of (\ref{HNCrepstat}).
The equation for $\tg$ is identical to the usual HNC equation
(\ref{HNC}), while $g_1$ is a solution of 
$g_1(x) = \exp(W_1(x))$, with $W_1$ given by the second equation
of (\ref{Wdef}) at $m=1$.

Within this static approach there is no obvious definition
of the dynamic energy. Previous work \cite{FraPar,Mon} suggests
the following computation:
 one considers a system at equilibrium at a 
temperature $T_R>T_G$ and an other system at temperature $T$. One defines 
the correlation functions
\bea
\rho^2 g_R(x-y) &=& <\rho(x)\rho(y)>_R, \nn
\rho^2 \tg(x-y) &=& <\rho(x)\rho(y)>,\nn
\rho^2 g_0(x-y) &=& <\rho(x)>_\al{(<\rho(y)>_R)}_\al, \nn
\rho^2 g_1(x-y) &=& <\rho(x)>_\al<\rho(y)>_\al,
\eea
where we have assumed that the states of the system at a given temperature are 
in correspondence with the states at lower temperature.  One imposes the 
constraint on the system at temperature $T$ that the function $g_0$ and $g_1$ 
are not zero.  If $T=T_R$ this constraint can be easily satisfied.  If $T<T_R$ 
this constraint forces the system at $T$ to stay out of equilibrium.  It has 
been conjectured in \cite{FraPar} that the correlation functions in the 
metastable state (from which we can easily extract the energy) are given by 
$g_1(x)$ computed at $T_R=T_M(T)$.

\section{Solution of the equations}

Let us first discuss some technical points which are common to the soft and the 
hard sphere cases.  In both cases the first task is to solve the replica 
symmetric HNC equation.  For spherically symmetric functions in dimension three 
we use the Fourier transform for the radial dependance, in the following form:
\be
q {\bf h}(q)   = 2 \pi  \int_0^\infty dr \sin(qr) r h(r).
\ee

We discretize this formula introducing in $r$ space a cutoff $R$ and a mesh size 
$a$.  In this way we have a simple formula for the inverse Fourier transform and 
we can also use the fast Fourier transform algorithm.  In most of the 
computations we have taken $a=1/32.5$ and $L=128*a \approx 4$.  We have tried 
smaller values of $a$ and larger values of $L$ without serious effects.  The 
solution of the equations can be found either by using a library minimization 
program \footnote {When one has to minimize the free energy with respect to one 
variable and maximize it with respect to another one, we first minimize with 
respect to the first variable, later maximize with respect to the second, and 
iterate the procedure until convergence.}, or a program which solves non linear 
equations.
 We have found first the solution at low enough density and then
followed it by continuity while gradually increasing the density.

\subsection{The soft sphere case}

The HNC equation gives a description of the liquid phase which is not perfect, 
but precise enough for our purpose: The energy (or equivalently the pressure), 
does not depart more than 15\% from the correct value (see fig.1), and the 
correlation function is also well reproduced (see fig.  2).

The numerical solution of the replicated HNC equations finds a dynamical 
transition at $\gamma\simeq 2.05$, and a static replica symmetry breaking 
solution at $\gamma \simeq 2.15$.  In numerical simulations the glass transition 
is found at a smaller value of $\gamma$, namely $\gamma=1.6$.  In the glass 
phase, the correlation function $g_1(r)$ is essentially a smoothed form of the 
function $\tg(r)$ plus an extra contribution at short distance which has 
integral near to $1$ (see Fig.3).  This form seems very reasonable: Considering 
the definition (\ref{gdef}) of $g_1$, we see that it basically characterizes the 
average over $\al$ of the product $\rho_\al(x)\rho_\al(y)$, which is precisely 
expected to have this kind of peak structure.

In spite of this nice form for $g_1$, this solution  has some
problems. A first one is found on the value of the
energy. The  
static 
 energy as function of $\gamma$ (or equivalently of $T$ at density $1$). 
 is plotted in fig.1. Although there is a discontinuity in 
 the specific heat, it is extremely small and  the final effects on the 
 internal energy are more or less invisible. The specific heat remains 
 extremely large. Moreover the value of $m$ has a very
unusual dependence on the 
 temperature (see fig.4). In all the known models with one 
step replica symmetry breaking, the breakpoint  $m$ 
 varies linearly with $T$  at low temperatures. 
Here we have a very different
behaviour.
 We have also computed the dynamical internal energy  
 and found out that it differs from the equilibrium one by an extremely 
 small account.
 
 We conclude that if we consider the qualitative behavior of the 
 correlation functions, we find a reasonable form, on the other hand the 
 energy in the glassy phase turns out to be quite wrong.

\subsection{The hard sphere case}

The situation is quite similar in the case of hard sphere.  The HNC 
approximation works reasonably in the liquid phase.  The pressure, which can be 
extracted either from the free energy or from its relation with $g(1^+)$, does 
not depart from the correct one by more than 15\% in the liquid phase.  The main 
defect is the absence of the peak around $r=\sqrt{3}$ and a too large value of 
the peak at $r=2$.  The pressure seems to diverge proportionally to 
$(\rho_c-\rho)^{-2}$, where $\rho_c$ is around 1.6, while the maximum possible 
density, corresponding to the fcc lattice, is $\sqrt{2}$.

Here replica symmetry is broken around $\rho\simeq1.19$ and the dynamical 
transition is located at $\rho\simeq1.17$.  These values are very close to the 
result of the numerical simulations which find a freezing transition around 
$\rho\simeq1.15$.  Unfortunately also in this case the computed value for the 
pressure differs by a very small amount from the replica symmetric one and it is 
therefore unacceptable.  Similar conclusions have also been reached for the 
Lennard Jones potential.

\section{Conclusions}

The simple implementation of the replica approach to glasses which we have 
proposed here provides some interesting results, like the existence of a glass 
transition at a reasonable value of the density.  However it is not 
satisfactory, in the sense that the effects of the transition on the 
thermodynamic of the systems are much too small.

At the moment we do not have a clear understanding of the reasons of this 
failure to grasp the thermodynamic properties of the glassy phase.  Two possible 
explanations came to our mind.  The first one would be that an approximation 
like that of HNC may miss some of the main physical characteristics of the 
problem in the glassy phase.  A first look at the corrections to the free energy 
indicates that they are quite large, lending some support to this hypothesis.  
More work is needed to decide on what class of diagrams should be added to cure 
this problem.  On the other hand we must also admit that we have not found so 
far a full derivation of the replicated HNC equations.  A more proper approach 
would be to work with a free energy expressed as a functional of the density 
$\rho(x)$, seeking all the (non translational invariant) solutions corresponding 
to glass phases.  Then the correlations and thermodynamics could be studied by 
giving to each solution a weight proportional to $\exp( -\beta F_\al)$, $F_\al$ 
being the free energy of the solution labeled by $\al$.

We think that finding an analytic approximation scheme which produces reasonable 
results in the glass phase is within reach.  The method we propose seems to be 
promising in this respect and shows how the replica method could be used to 
study the glass phase.

\section*{Aknowledgements} We thank  D. Dean and R. Monasson for useful discussions.

\section*{Figure Captions}

{\bf Fig.1} The static energy of a system of soft spheres 
as a function of the dimensionless inverse density $\gamma$ from
 numerical simulation (squares), from
 the replica symmetric HNC equation (full line) and from one step
replica symmetry broken 
 HNC equation (open circles). 

\noindent{\bf Fig.2} 
 The correlation function $\tg(r)$ of a system of soft spheres 
as a function of the distance at the dimensionless  density
 $\gamma =1.6$ corresponding to  the numerical glassy 
 transition:
  numerical simulations (points) and replica symmetric HNC 
 equation (full line).

\noindent{\bf Fig.3} 
 The correlations  $\tg(r)$  (full line)
 and $g_1(r)$ (broken line) obtained from the replicated hnc equations,
as  functions of the distance, for soft spheres at the density where 
 replica symmetry is broken ($\gamma=2.15$).

\noindent{\bf Fig.4} 
 The value of the breakpoint $m$ in the replica symmetry breaking
solution for soft spheres as a function of 
the dimensionless density $\gamma$.


\begin{thebibliography}{99}
\bibitem{Science} For a recent introduction, see the papers in Science,
{\bf 267} (1995) 1924, and the lectures in  {\it Liquids, freezing and glass transition},
 Les Houches 1989, JP Hansen, D. Levesque, J. Zinn-Justin Editors,  North Holland.
\bi{KiThi}
T. R. Kirkpatrick and D. Thirumalai, Phys. Rev. Lett. {\bf 58},
2091 (1987); Phys. Rev. {\bf B36}, 5388 (1987),
 J. Phys. {\bf A22}  L149(1989). 

\bi{KiThiWo}
T. R. Kirkpatrick, D. Thirumalai and P.G. Wolynes,  Phys. Rev. {\bf
A40}, 1045 (1989)
\bi{Parglass}
G.Parisi, {\it Slow dynamics in glasses}, cond-mat {\bf 941115} and {\bf 9412034}.

\bi{FrHe}
S. Franz and J. Hertz; Phys. Rev. Lett. {\bf 74} (1995) 2114.

\bi{BCKM}
J.P.Bouchaud, L.Cugliandolo, J. Kurchan and M.M\'ezard,
"Mode coupling approximations, glass theory and disordered systems",
 preprint condmat 9511042, to appear in Physica A.

\bi{BM}
J.-P. Bouchaud and M. M\'ezard; J. Physique I (France) {\bf 
4} (1994) 1109.

\bi{MPR}
E.  Marinari, G.  Parisi and F.  Ritort; J.  Phys.  {\bf A27} (1994) 7615; J.  
Phys.  {\bf A27} (1994) 7647.

\bi{SI3}
L. F. Cugliandolo,  J. Kurchan, G. Parisi and F.Ritort;
Phys. Rev. Lett. {\bf 74} (1995) 1012. 

\bi{Joseph}
P. Chandra, L. Ioffe and D. Sherrington; cond-mat {\bf 9502018}.
P. Chandra, M. Feigelmann and L. Ioffe;
preprint cond-mat  {\bf 9509022}.

\bi{JosephPar}
G. Parisi, , {\sl $D$-dimensional Arrays of Josephson Junctions, Spin
Glasses and  $q$-deformed Harmonic  Oscillators}, cond-mat preprint (1994).

\bi{MPR_FF}
 E. Marinari, G. Parisi and F. Ritort,  {\sl Fully Frustrated
Ising Spin Model on the Hypercube is Glassy and Aging}, cond-mat preprint 
(1994).

\bi{CKMP}
L.F. Cugliandolo, J. Kurchan, R. Monasson and G. Parisi,
{\sl A mean field hard-spheres model of glass}, J. Phys. (in press).
\bi{CriHorSom}
A. Crisanti, H. Horner and H.-J. Sommers,  Z. Phys. {\bf
B92}  257 (1993).

\bi{CuKu}
 L. F. Cugliandolo and J.Kurchan, Phys. Rev. Lett. {\bf 71}, 1
(1993).

\bi{FraPar}
S. Franz and G. Parisi, J.Phys. I (France) {\bf 5} (1995) 1401.

\bi{Mon}
R. Monasson, Phys.Rev.Lett. {\bf 75} (1995) 2847.

\bi{BaBuMez}
A.Barrat, R.Burioni and M.M\'ezard, "Dynamics within metastable
states in simple spin glass systems", preprint cond-mat/9511089.

\bi{hiw} Y. Hiwatari, J. Phys. {\bf C13} (1980) 5899.

 \bibitem{MPV}  M.~M\'ezard, G.~Parisi and  M.~A.~Virasoro,  {\em
Spin Glass Theory  and Beyond}, World Scientific, (Singapore 1987).

\bi{HanMc}
See for instance
J.P. Hansen and I.R. Macdonald, "Theory of simple liquids",
(Academic, London, 1986), or H.N.V. Temperley, J.S. Rowlinson and
G.S. Rushbrooke, "Physics of simple liquids", NorthHolland (Amsterdam 1968).

\bi{RamYus}
T.V.Ramakrishnan and M.Yussouf, Phys.Rev. {\bf B19} 2775(1979).

 \end{thebibliography}
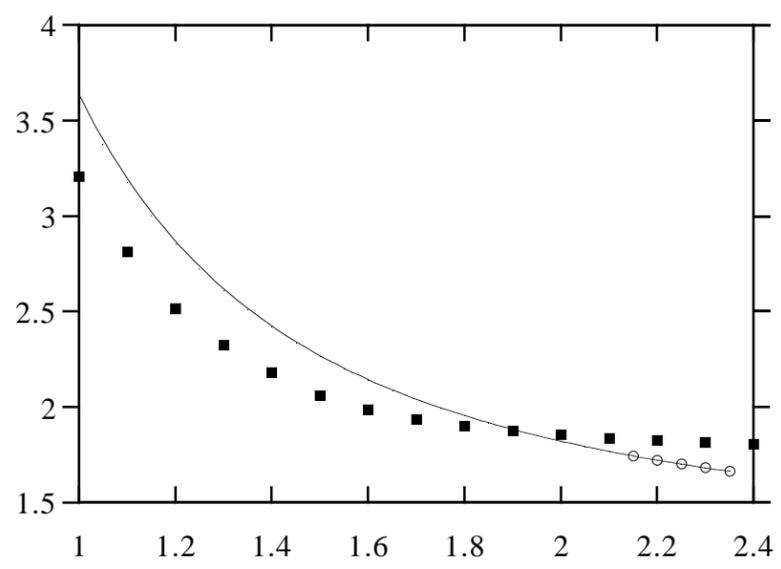
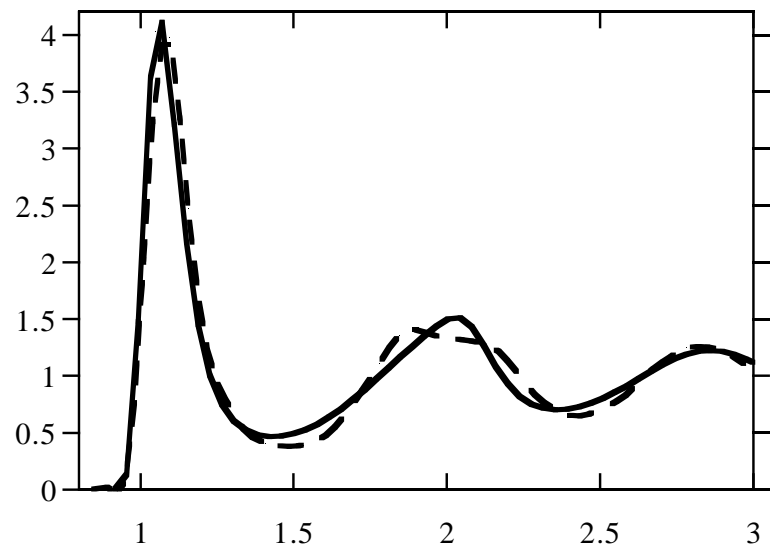
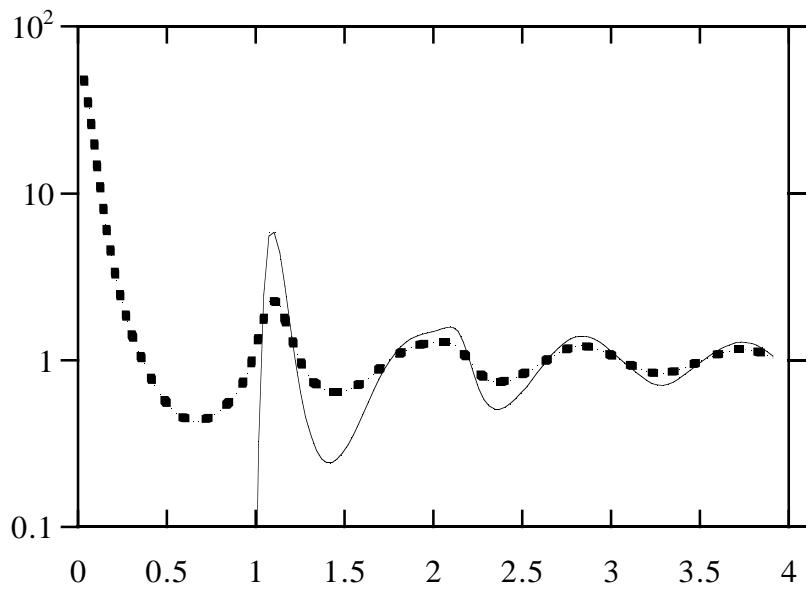
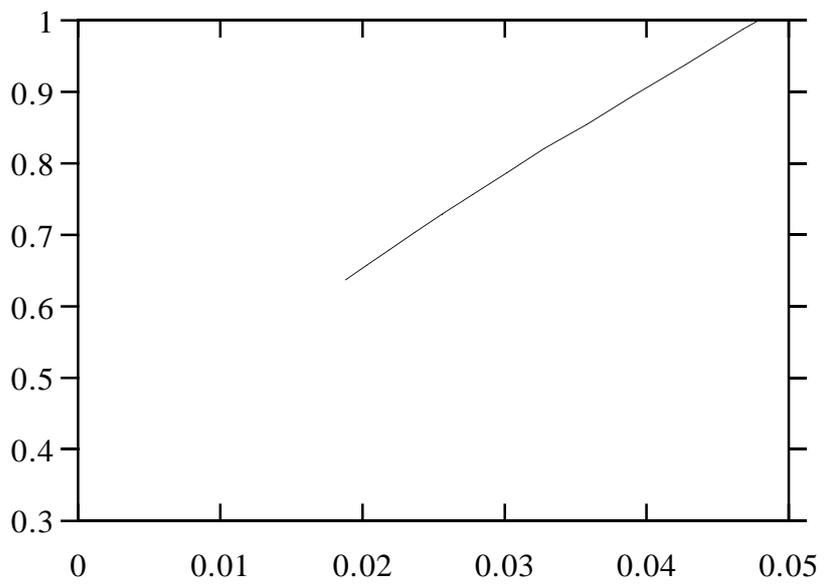
\end{document}